\documentclass[10pt]{article}

\usepackage[english]{babel}

\usepackage{graphics,color}
\usepackage{newlfont,epsfig}

\begin{document}
\date{}
\begin{center}
{\Large\bf Deviations from reversible dynamics in a qubit-oscillator system coupled to a very small environment}
\end{center}
\begin{center}
{\normalsize A. Vidiella-Barranco \footnote{vidiella@ifi.unicamp.br}}
\end{center}
\begin{center}
{\normalsize{ Instituto de F\'\i sica ``Gleb Wataghin'' - Universidade Estadual de Campinas}}\\
{\normalsize{ 13083-859   Campinas  SP  Brazil}}\\
\end{center}
\begin{abstract}
In this contribution it is considered a simple and solvable model consisting of a qubit in interaction with an oscillator
exposed to a very small ``environment" (a second qubit). An isolated qubit-oscillator system having the oscillator initially 
in one of its energy eigenstates exhibits Rabi oscilations, an evidence of coherent quantum behaviour. It is shown here in which 
way the coupling to a small ``environment" disrupts such regular behaviour, leading to a quasi-periodic dynamics for the qubit 
linear entropy. In particular, it is found that the linear entropy is very sensitive to the amount of mixedness of 
the ``environment". For completeness, fluctuations in the oscillator energy are also taken into account.

\end{abstract}

% \linenumbers
\setcounter{footnote}{0}
\section{Introduction}

Exactly solvable models often give new insights into the properties of simple quantum systems. A well known of 
such models is the Jaynes-Cummings model \cite{jay63} of quantum optics: a two-level system (qubit) coupled to the quantized field
(oscillator) under the rotating wave approximation (RWA). A peculiar feature of that model is the periodic exchange of excitations between 
the qubit and the oscillator, namely Rabi oscillations of the qubit population, if the oscillator is initially prepared in a
number state $|n\rangle$, an energy eigenstate of the free oscillator. A similar behaviour is found for other properties of the 
qubit, such as the von Neumann entropy, for instance, a quantifyer of the purity of the qubit state. Perfect Rabi oscillations of the 
entropy mean that a system initially in a pure state is able to return to a pure state at a later time, and they are therefore a 
signature of coherent quantum behaviour.
On the other hand, if the oscillator is initially prepared in a distribution of number states, e.g., a coherent state, 
entropy will be no longer a periodic function of time, and the qubit will not be able to return to a pure state \cite{kni88,len98}. 
Thus, if the qubit is coupled to a quantum oscillator in a state which is itself a superposition of number states, the qubit 
evolution, in particular, its entropy, will become significantly disordered. 
This is because the incommensurate Rabi frequencies $\Omega \propto \sqrt{n+1}$ present in the Jaynes-Cummings solution introduce
dephasing, leading to an irreversible-like behaviour. 
Another possible source of disorder, even for the oscillator initially in a single number state $|n\rangle$ is related to chaos, 
and arises in a strongly coupled qubit-oscillator system without making the RWA \cite{grigo91}. 

In addition to that, the investigation of the evolution of simple quantum systems coupled to an environment has been a subject 
of continuing interest for many years \cite{zeh70}. In very precise experiments, it has been found that the Rabi oscillations in a Jaynes-Cummings
system are indeed irreversibly damped \cite{haro96}. This important phenomenon is attributed to external influences (noise), which are normally
modelled by coupling the system of interest to a very large number of other quantum systems (reservoir). 
Those phenomenological models \cite{cald85}, based on the 
assumption that a large system having certain pre-defined properties is coupled to a subsystem of interest, are in general
able to describe the main features of dissipative quantum subsystems: loss of energy and loss of quantum coherence.
However, rather than being coupled to a large reservoir, a quantum system may be subjected to an uncontrollable coupling to external 
systems having few  degrees of freedom, and one may ask in which ways such a small ``environment'' may disturb the quantum features of the
system of interest.  
In this paper I am going to consider a model consisting of a two-level system (qubit1) interacting with an oscillator weakly 
coupled to a single sub-system constituted by a second two-level system (qubit2) which will in fact play the role of a small 
``environment''. This corresponds to another well known model of quantum optics: the two-atom Tavis-Cummings model 
\cite{tav68,zub87}. Here, though, I will explore this symple quantum system from a different perspective: the qubit1-oscillator
will be our system of interest, and qubit2 will constitute a perturbing environment.
One of the advantages of employing the Tavis-Cummings model, is that it is exactly solvable under the RWA, 
which will allow a general, non-perturbative treatment of the dynamics of qubit1. The main purpose of this contribution is to find 
in which way quantum reversibility represented by the Rabi oscillations of entropy in the number-state Jaynes Cummings model could 
be affected by coupling the oscillator to a small system and how that disturbance depends on the features of such an ``environment". 
In particular, I am going to analyze the dynamics of the linear entropy of qubit1 for different degrees of mixedness of the ``environment" 
(qubit2). I would like to point out that it is not my intention to present a model to reproduce a phenomenon such as irreversible 
loss of quantum coherence, but rather, I am interested in examining the deviations from perfect reversibility due to a (unwanted) 
coupling to a ``small environment".

The paper is organized as follows. In section 2, I will present a preliminary discussion based on the Jaynes-Cummings model for simple
preparations of the oscillator (pure and mixed). In section 3, I will show the solution of the actual system-``environment" model and 
the calculation of the linear entropy of qubit1. In section 4, I will discuss the time-evolution of the linear entropy for different initial 
conditions as well as the influence of energy fluctuations in the oscillator. In section 5 I will summarize the conclusions.

\section{Jaynes-Cummings model with an initial mixed state\label{sec2}}

\begin{figure}[h]
	\centering
	\resizebox{0.5\columnwidth}{!}{
		\includegraphics{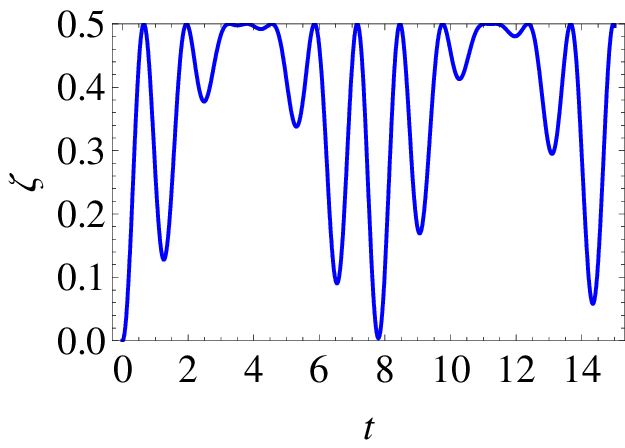}}
	\caption{\label{figure1} Plot of the linear entropy $\zeta$ (as a function of $t$) of a qubit initially in state $|e\rangle$ and the 
	oscillator initially in the mixed state $\rho_{osc}(0) = f |0\rangle\langle 0| + (1-f) |1\rangle\langle 1|$ with $\lambda = 1.0$ and $f = 0.5$.}
\end{figure}

In the ``pure state" Jaynes-Cummings model, if the oscillator is initially prepared in a number state, the entropy has a regular behaviour;
perfect Rabi oscillations, as a single frequency is involved. However, if the oscillator is initially in a mixed state, different excitation 
numbers will come into play, yielding different Rabi frequencies. A simple situation considering a small number of excitations for the oscillator 
could be an initial mixed state involving just the vacuum $|0\rangle$ and the first excited state $|1\rangle$
\begin{equation}
\rho_{osc}(0) = f |0\rangle\langle 0| + (1-f) |1\rangle\langle 1|.
\end{equation}
Here qubit1 (with states $|e\rangle$ and $|g\rangle$) is assumed to be initially in the excited state $|e\rangle$. 
The qubit1-oscillator joint density operator (Jaynes-Cummings) may then be written as
\begin{eqnarray}
\rho(t) &=& \hat{A}(t) \rho_{o}(0) \hat{A}^\dagger(t) |g\rangle\langle g| + \hat{A}(t) \rho_{o}(0) \hat{B}^\dagger(t) |g\rangle\langle e| \nonumber \\
&+& \hat{B}(t) \rho_{o}(0) \hat{A}^\dagger(t) |e\rangle\langle g| + \hat{B}(t) \rho_{o}(0) \hat{B}^\dagger(t) |e\rangle\langle e|,
\end{eqnarray}
where $\hat{A}(t) = \cos(\lambda t \sqrt{\hat{a} \hat{a}^\dagger})$ and 
$\hat{B}(t) = -i\hat{a}^\dagger\sin(\lambda t \sqrt{\hat{a} \hat{a}^\dagger})/\sqrt{\hat{a} \hat{a}^\dagger}$. Here, $a$ ($a^{\dagger}$) are the oscillator 
annihilation (creation) operators and $\lambda$ is the qubit-oscillator coupling constant.
After tracing over the oscillator's degrees of freedom, one obtains the reduced density 
operator for qubit1, $\rho_{q1}(t) = Tr_{o}\left[\rho(t)\right]$. Then, a straightforward calculation gives us the linear 
entropy\footnote{The linear entropy is zero for a pure state ($Tr\rho^2 = Tr\rho = 1$), and 
different from zero for a statistical mixture. In our specific case, the linear entropy varies from 0 to 0.5.} for qubit1,
$\zeta(t) = 1 - Tr\left[\rho^2_{q1}(t)\right]$, or
\begin{eqnarray}
\zeta(t) &=& 1 - \big[ f\cos^2(\lambda t) + (1-f) \cos^2(\sqrt{2}\lambda t)\big]^2 \nonumber \\
& - & \big[ f\sin^2(\lambda t) + (1-f) \sin^2(\sqrt{2}\lambda t)\big]^2.
\end{eqnarray}
That is an example of a ``quasi periodic function", a sum of periodic functions with incommensurate frequencies $\nu_1=\lambda$ and 
$\nu_2=\sqrt{2}\lambda$. In figure (\ref{figure1}) it is plotted $\zeta$ as a function of time for $f = 0.5$. 
We note a damping of the oscillations, followed by an increase of of their 
amplitude so that $\zeta$ gets very close to zero again after a few cycles. Stricktly speaking, though, qubit1 does not return to a pure 
state, although $\zeta$ may get very close to zero in this case. This simple example will be useful to  
understand the behaviour of the qubit1-oscillator dynamics coupled to an external system, as follows.

\section{The model}

The model considered here consists of a two-level system, qubit1, coupled to a quantized field mode (oscillator), which is itself 
coupled to a second two-level system: qubit2, the ``environment''. The hamiltonian describing such a system (within the RWA) 
is the familiar two-atom Tavis-Cummings hamiltonian \cite{tav68,zub87}, or
\begin{equation}
  H = \hbar\omega_o a^\dagger a + \frac{\hbar}{2}\sum_{i=1}^{2} \omega_{i,q} \sigma_{i}^z + \hbar\sum_{i=1}^{2}\lambda_i\left(a\sigma_i^+ 
  + a^\dagger\sigma_i^- \right),
\end{equation}
where $a$ ($a^{\dagger}$) are the oscillator annihilation (creation) operators, and $\sigma_i^z,\! \sigma_i^+,\! \sigma_i^-$ are the Pauli operators for the 
for the $i$-th qubit. I am going to consider the fully resonant case, i.e., having both qubit1 and qubit2 frequencies equal to the oscillator frequency, 
$\omega_{i,q} = \omega_o$. However, the oscillator-qubits couplings, $\lambda_1$ and $\lambda_2$ will be assumed to be different, so that it will be 
straightforward to simulate a weakly coupled subsystem. That model has a well known exact analytical solution 
\cite{zub87}, but here I am going to use initial conditions different than the usual ones: while qubit1 and the oscillator will be assumed to be initially 
prepared in pure states, qubit2, the ``environment", will be in a more general, mixed state. The initial state of the whole system will be given by 
the product state $\rho(0) = \rho_1(0)\otimes\rho_{osc}(0)\otimes\rho_2(0)$, where
\begin{equation}
 	\rho_1(0) = |e_1\rangle\langle e_1|, \ \ \ \ \rho_{osc}\left(0\right) = |\phi\rangle\langle\phi|, \ \ \ \ 
 	\rho_2(0) = p|e_2\rangle\langle e_2| + (1-p)|g_2\rangle\langle g_2|,
\end{equation}
or qubit1 in the upper state, the oscillator in a generic pure state 
$|\phi\rangle = \sum B_n |n\rangle$, (with $B_n$ real) and qubit2 in a statistical mixture of upper and 
lower states. The time-evolved system density operator may be expressed as a sum of two terms
\begin{equation}
	\rho(t) = p \left|\Psi_p (t)\right\rangle \left\langle \Psi_p (t)\right| +
	          (1-p) \left|\Psi_{1-p} (t)\right\rangle \left\langle \Psi_{1-p} (t)\right|,
\end{equation}
where
\begin{eqnarray}
	\left|\Psi_p(t)\right\rangle = \sum_n B_n \Big[ C_1(t)\left|e_1,e_2,n\right\rangle + C_2(t) \left|e_1,g_2,n+1\right\rangle + \nonumber\\ 
	 + C_3(t) \left|g_1,e_2,n+1\right\rangle + C_4(t) \left|g_1,g_2,n+2\right\rangle \Big],\label{psie}
\end{eqnarray}
and
\begin{eqnarray}
	\left|\Psi_{1-p}(t)\right\rangle = \sum_n B_n \Big[ C'_1(t)\left|e_1,e_2,n-1\right\rangle + C'_2(t) \left|e_1,g_2,n\right\rangle + \nonumber\\ 
	 + C'_3(t) \left|g_1,e_2,n\right\rangle + C'_4(t) \left|g_1,g_2,n+1\right\rangle \Big].\label{psig}
\end{eqnarray}
After solving the Schr\"odinger equation, we obtain the coefficients $C(C')$, which are given by 
\begin{eqnarray}
	C_1(t) &=& \left(\frac{1}{2D}\right)a_- \cos d_+t +a_+ \cos d_-t \nonumber \\
	C_2(t) &=& -\left(\frac{i}{2D}\right)\left( a_-b_+\right)^{(1/2)}\sin d_+t - \left( a_+b_-\right)^{(1/2)}\sin d_-t \nonumber \\
	C_3(t) &=& -\left(\frac{i}{2D}\right)\left( a_-b_-\right)^{(1/2)}\sin d_+t - \left( a_+b_+\right)^{(1/2)}\sin d_-t \nonumber \\
	C_4(t) &=& \left(\frac{1}{2D}\right)\left( a_+a_-\right)^{(1/2)}\left[\cos d_+t - \cos d_-t \right],\label{ces}
\end{eqnarray}
where $a_\pm = D \pm (\lambda_1^2+\lambda_2^2)$, 
$b_\pm = D \pm (\lambda_1^2-\lambda_2^2)$, with
\begin{equation}
D = \left[(2n+3)^2(\lambda_1^2+\lambda_2^2)^2-4(n+1)(n+2)(\lambda_1^2-\lambda_2^2)^2\right]^{1/2},\label{Dcapital}
\end{equation}
\begin{equation}
d_\pm = \left[(2n+3)(\lambda_1 + \lambda_2)^2 \pm D\right]^{1/2}/\sqrt{2}.\label{dplusminus}
\end{equation}
Also
\begin{eqnarray}
	C'_1(t) &=& -\left(\frac{i}{2D'}\right)\left( a'_-b'_+\right)^{(1/2)}\sin d'_+t - \left( a'_+b'_-\right)^{(1/2)}\sin d'_-t \nonumber \\
	C'_2(t) &=& \left(\frac{1}{2D'}\right)b'_- \cos d'_+t - b'_+ \cos d'_-t \nonumber \\
	C'_3(t) &=& \left(\frac{1}{2D'}\right)\left( b'_+ b'_-\right)^{(1/2)}\left[\cos d'_+t - \cos d'_-t\right] \nonumber \\
	C'_4(t) &=& -\left(\frac{i}{2D'}\right)\left( a'_+ b'_+\right)^{(1/2)}\sin d'_+t + \left( a'_- b'_-\right)^{(1/2)}\sin d'_-t.\label{cesprime}
\end{eqnarray}
The ``primes" in the expressions above ($D'$, etc.) mean that the corresponding terms are evaluated by replacing $n$ by $n-1$. 

I am primarily interested in investigating the influence of the environment (qubit2) on the dynamics of qubit1. 
For instance, in order to find deviations from reversible behaviour in the qubit1-oscillator interaction, we may calculate 
the linear entropy relative to qubit1, obtained from its reduced density operator by tracing over qubit2 and the oscillator variables, or 
$\rho_{q1}(t) = Tr_{q2,o}\left[\rho(t)\right]$. After some algebra, we obtain
\begin{equation}
	\zeta = 1 - Tr\left[\rho^2_{q1}(t)\right] = 1 - \left( \alpha^2 + \beta^2 + 2|\gamma|^2 \right),\label{zetaex}
\end{equation}
where
\begin{equation}
	\alpha = \sum_n (B_n)^2\left[p\left(|C_{3,n}|^2 + |C_{4,n}|^2\right) + (1-p)\left(|C'_{3,n}|^2 + |C'_{4,n}|^2\right) \right],\label{alpha}
\end{equation}

\begin{equation}
	\beta = \sum_n (B_n)^2\left[p\left(|C_{1,n}|^2 + |C_{2,n}|^2\right) + (1-p)\left(|C'_{1,n}|^2 + |C'_{2,n}|^2\right) \right],\label{beta}
\end{equation}
and
\begin{eqnarray}
	\gamma &=& \sum_n B_n B_{n+1} \Big\{p\left[C_{4,n}C^*_{2,n+1} + C_{3,n}C^*_{1,n+1}\right] + \nonumber \\
	&+& (1-p) \left[C'_{4,n}C'^*_{2,n+1} + C'_{3,n}C'^*_{1,n+1}\right] \Big\}.\label{gamma}
\end{eqnarray}

\section{Linear entropy - analysis and numerical results}

\subsection{Oscillator initially prepared in a number state $|N\rangle$} 

Now I am going to to analyse the time evolution of the linear entropy of qubit1 for different conditions of the oscillator and 
environment. I would like to recall that we are actually dealing here with systems having a few
degrees of freedom, in contrast to the usual theory of open quantum systems, in which the relevant subsystems are assumed to be coupled to
reservoirs with a large number of degrees of freedom. Our ``environment" is constituted by one of the smallest possible disturbing systems: 
a single two-level system. As we have already seen, in the absence of an environment on any kind, if the oscillator is initially prepared 
in a number state (oscillator energy eigenstate), the linear entropy of qubit1 will be simply a periodic function of time, i.e., it will 
perform Rabi oscillations. This makes evident the perfectly reversible character of the qubit1-oscillator dynamics in such an ideal case. 
Now, what would be the influence of the small environment (qubit2) on the time evolution of the linear entropy of qubit1? Next I am going
to analyse the behaviour of the linear entropy for different degrees of mixedness of the environment. The oscillator will be initially 
prepared in a number state $|N\rangle$, or $B_n = \delta_{n,N}$.

\begin{figure}[h]
	\centering
	\resizebox{0.5\columnwidth}{!}{
		\includegraphics{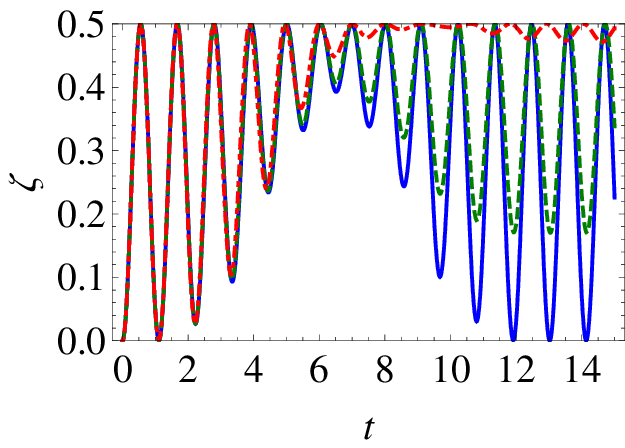}}
	\caption{\label{figure2} Plot of the linear entropy $\zeta$ (as a function of $t$) of qubit1 initially in state $|e\rangle$, the oscillator initially in
	the number state $|1\rangle$ and qubit2 initially: a) in a pure state, $p = 0.0$ (blue, continuous curve); b) in a mixed state, $p = 0.1$ 
	(green, dashed curve); c) in a maximally mixed state, $p = 0.5$ (red, dot-dashed curve). In all cases $\lambda_1 = 1.0$ and $\lambda_2 = 0.1$.}
\end{figure}

Firstly, I would like to point out the peculiar structure of equation (\ref{dplusminus}); $d_+$ and $d_-$ are basically the
relevant frequencies of the dynamics, as the linear entropy contains terms oscillating at them. A very important point is that 
$d_+$ and $d_-$ are in general incommensurate. Of course if the oscillator is completely decoupled from   
qubit2 ($\lambda_2 = 0.0$) we will have $d_+ = d_-$ and the linear entropy of qubit1 will be a truly periodic function. 
Otherwise, if we ``turn on" the environment, the linear entropy will become a quasi periodic function of time, instead. 
We now assume a relatively weak coupling; for instance, by choosing $\lambda_2 = 0.1$ and $\lambda_1 = 10\lambda_2$. 
We may first consider the environment initially in a pure state ($p = 0.0$). In this case frequencies $d_+$ and $d_-$ are different,
the oscillations will be modulated, and we have again a quasi periodic function.
In figure (\ref{figure2}a) we have a plot of the linear entropy as a function of time in that situation. 
The linear entropy increases but almost returns to zero after a few cycles, similarly to the case 
we have seen in section \ref{sec2}, i.e., the oscillator initially in mixed state [see figure (\ref{figure1})] but without the
presence of an ``environment". This means that the effect (on qubit1) of an ``environment" of this type (qubit2) initially in a pure state,
may have some resemblance to the case in which there is no environment but the oscillator is initially in a mixed state. 

\begin{figure}[ht]
	\centering
	\resizebox{0.5\columnwidth}{!}{
		\includegraphics{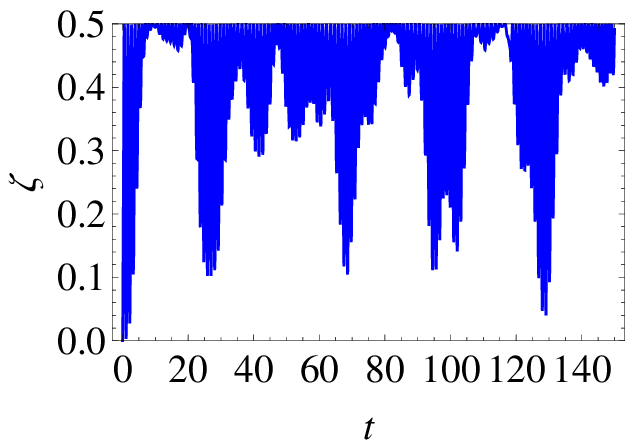}}
	\caption{\label{figure3} Plot of the linear entropy $\zeta$ (as a function of $t$) of qubit1 initially in state $|e\rangle$, the oscillator initially in
	the number state $|1\rangle$ and qubit2 initially in a maximally mixed state, $p = 0.5$ for a longer time-scale, with $\lambda_1 = 1.0$ and $\lambda_2 = 0.1$}
\end{figure}

However, the situation is quite different if the environment is initially in a mixed state; in this case terms containing two more 
different (and incommensurate) frequencies, $d'_+$ and $d'_-$, are introduced; see expressions (\ref{zetaex}) to (\ref{gamma}). 
We expect, then, that this will cause a stronger dephasing. In fact, even a small amount of ``mixedness" in qubit2 will bring a significant 
disorder to qubit1, as shown in figure (\ref{figure2}b), where qubit2 is in a mixed state with $p = 0.1$. Of course the effect of the ``environment"
is even more destructive if qubit2 is initially in a maximally mixed state, i.e., (p = 0.5), as shown in figure (2c). 
In the discussion above a relatively short time-scale has been considered, 
but as a quasi periodic function is involved, we expect some sort of recurrence at longer times. Nevertheless, even for a larger time-scale, 
we note a very irregular evolution of $\zeta$, indicating that the system spends most of the time in a highly mixed state, as shown in 
figure (\ref{figure3}). As we would expect, a noisier environment (maximally mixed state) has a much stronger impact 
on qubit1 linear entropy $\zeta$ compared to a pure state environment case. Notwithstanding, this model does not lead to 
full irreversibility in the sense that the entropy grows to a maximum value for long times. Moreover, dissipation does not take place, as
excitations go back and forth between system and environment.

\subsection{Oscillator prepared in a binomial state: effect of energy fluctuations}

Now suppose that in addition to the coupling of the oscillator to an external system (qubit2), there is some noise due
to imperfections in the preparation of the state's oscillator, e.g., energy fluctuations. In order to take into account such effects, 
I will consider the oscillator initiallly prepared in a binomial state\footnote{The binomial state is characterized by two independent parameters: 
the maximum permissible number of excitations, $M$, and the probability of having a single excitation, $q$.} \cite{tei85}, 
with coefficients $B_n$ given by
\begin{equation}
B_n^M = \left[\frac{M!}{(M-n)!n!}q^n(1-q)^{M-n}\right]^{1/2}.
\label{eq:binomialcoef}
\end{equation}
Therefore, in equations (\ref{psie}) and (\ref{psig}) the summations are up to the finite number $M$, i.e., $|\phi\rangle = \sum_{n=0}^M B_n |n\rangle$.
In equations (\ref{alpha}) and (\ref{beta}) the summations run up to $n=M$ and in (\ref{gamma}) to $n=M-1$.
The binomial states have been already considered in the context of the two-atom Tavis-Cummings model \cite{sha86}, but the authors were concerned with the
dynamics of atomic populations and photon number distributions, instead. Binomial states are in fact very convenient for my purposes mainly 
because of their interpolating properties: if the parameter $q=1$, the binomial state becomes the number state $|M\rangle$; if $q\rightarrow 0$ and 
$M\rightarrow\infty$, but having $Mq = \alpha^2$ finite, the binomial state becomes the coherent state $|\alpha\rangle$. The energy fluctuations may 
be quantified via the Fano factor, defined as the dispersion of excitation number divided by the mean excitation number, or 
$F = \left\langle \Delta\hat{n}^2\right\rangle/\left\langle \hat{n}\right\rangle$. For a binomial state, $F = 1-q$. Note that $F = 0$ for a 
number state. First, I would like to make a preliminary discussion about the effect of fluctuations without having the oscillator coupled to qubit2.

\subsubsection{Absence of an environment}

\begin{figure}[ht]
	\centering
	\resizebox{0.5\columnwidth}{!}{
		\includegraphics{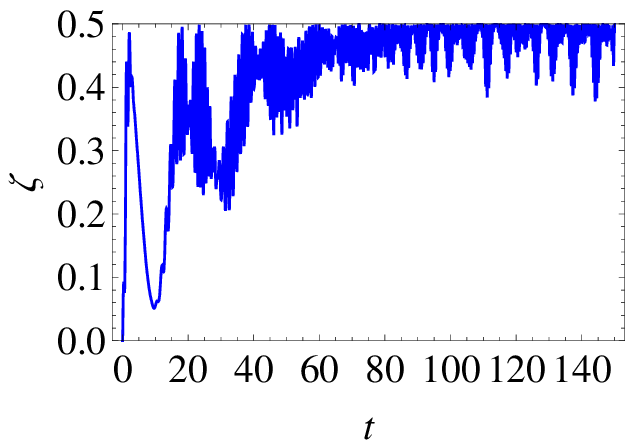}}
	\caption{\label{figure4} Plot of the linear entropy $\zeta$ (as a function of $t$) of qubit1 initially in state $|e\rangle$ and the oscillator 
	initially in a binomial state with $M = 100$ and $q = 0.1$. In this case qubit2 is decoupled, $\lambda_2 = 0.0$ and $\lambda_1 = 1.0$.}
\end{figure}

It is well known that if the oscillator is initially prepared in a distribution of number states, rather than in number states, 
the von Neumann entropy (or the linear entropy) of the qubit will have a very different behaviour.  
For instance, if the oscillator is initially prepared in a (pure) coherent state, the qubit1 entropy will sharply increase 
to a maximum value and then decrease to a minimum that may be very close to zero
\cite{kni88,len98}; a larger initial oscillator mean energy will bring the oscillator closer to a pure state at a specific time.  
As time goes on, the entropy grows again and starts showing an irregular behaviour. A very similar behaviour is shown in figure 
(\ref{figure4}), for an initial binomial state having $M = 100$ and $q = 0.1$, i.e., a good approximation for a coherent state.
In this case the entropy oscillates but it will grow progressively, showing that the system will remain in a mixed state even at 
very long times \cite{len98} (irreversible-like behaviour). 

\begin{figure}[h]
	\centering
	\resizebox{0.5\columnwidth}{!}{
		\includegraphics{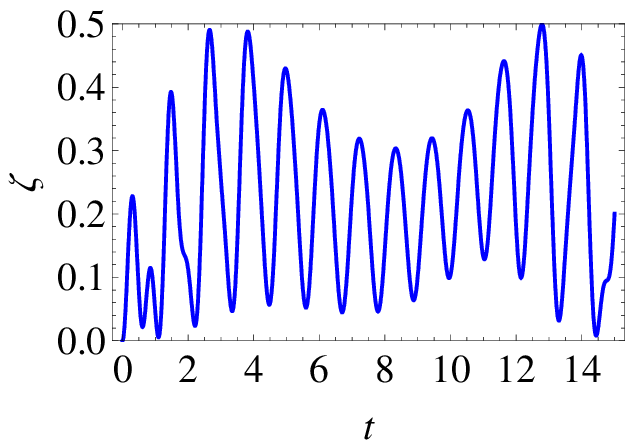}}
	\caption{\label{figure5} Plot of the linear entropy $\zeta$ (as a function of $t$) of qubit1 initially in state $|e\rangle$ and 
	the oscillator initially in a binomial state with $M = 7$ and $q = 0.85$. In this case qubit2 is decoupled, $\lambda_2 = 0.0$ and $\lambda_1 = 1.0$.}
\end{figure}

As an example of deviations from coherent oscillations characteristic of the number state case, we have in figure (\ref{figure5}) 
the short time behaviour of the linear entropy for qubit1 in the case of a binomial state with $M = 7$ and $q = 0.85$. 
In this case the linear entropy also oscillates and at a later it gets very close to zero. Therefore, a small energy spread in the 
initial oscillator state may also induce a quasi-periodic behavior to the qubit1 linear entropy.

\subsubsection{Presence of an environment}

\begin{figure}[h]
	\centering
	\resizebox{0.5\columnwidth}{!}{
		\includegraphics{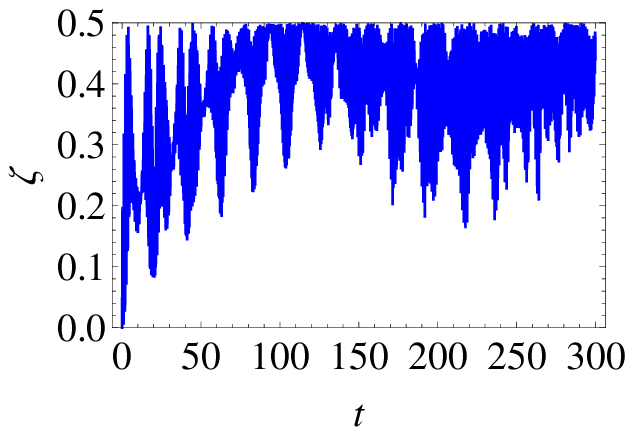}}
	\caption{\label{figure6} Plot of the linear entropy $\zeta$ (as a function of $t$) of qubit1 initially in state $|e\rangle$ and 
	the oscillator initially in a binomial state with $M = 11$ and $q = 0.9$. In this case qubit2 is coupled to the oscillator, with 
	$\lambda_2 = 0.1$, $\lambda_1 = 1.0$, and $p = 0.5$.}
\end{figure}

Now I would like to discuss the combined effects of fluctuations in excitations and the environment, as both favour irregular evolution of
the linear entropy. Suppose that the oscillator has a relatively small energy spread, e.g., having Fano factor $F = 0.1$, relatively 
close to a number state, and it is coupled to a maximally mixed ``environment", $p = 0.5$. We note a significant departure from a
regular oscillatory behaviour. Thus, the joint influence of the oscillator energy fluctuations and the weakly coupled ``environment" (qubit2) 
is quite apparent even for longer times, as shown in figure (\ref{figure6}).  

\section{Conclusion}

In this work, I have considered the time evolution of the linear entropy of a two-level system (qubit1) coupled to an oscillator when the latter is 
interacting with a noisy two-level system (qubit2). This simple model has analytical solution within the RWA, which allowed a non-perturbative 
treatment and the accurate investigation of the dynamics of the system at all times. Nevertheless, despite being an integrable system, a relatively 
complex dynamics may arise; this is of course related to peculiar features of the dynamics, the linear entropy containing terms with frequencies 
that are in general incommensurate, which results in a quasi periodic dynamics. Besides, uncertainties in the ``environment" (quantified by its 
degree of mixedness) have a significant influence on the qubit1 dynamics. In particular, the evolution of the linear entropy is 
highly sensitive to variations in the parameter $p$, or the initial mixedness of qubit2. Therefore, despite of the fact that the 
system under consideration is non-chaotic, it might show a (quasi periodic) 
very irregular evolution. I should recall that the situation is
different from the one in which qubit2 is not coupled ($\lambda_2 = 0.0$) and the oscillator
is initially in a mixed state, as discussed in section 2. There are a few
similarities, but the presence of a third subsystem (qubit2) brings about, for
instance, coherent exchange of excitations between the system and its “environment”,
and a more complex dynamics. I have also considered the effect
of energy fluctuations in the initial preparation of the oscillator; even a small
spread in energy will enhance the irregular behaviour of the linear entropy.
Thus, the reversibility represented by the perfect periodic behaviour of Rabi
oscillations may be strongly affected by a small “environment”, specially if
there are additional sources of noise involved.  



%% References


%\bibliographystyle{elsarticle-num}



\section*{Acknowledgements}

I would like to thank CNPq (Conselho Nacional para o 
Desenvolvimento Cient\'\i fico e Tecnol\'ogico) and FAPESP
(Funda\c c\~ao de Amparo \`a Pesquisa do Estado de S\~ao Paulo), Brazil,
for financial support through the National Institute for Science and 
Technology of Quantum Information (INCT-IQ) and the Optics and Photonics 
Research Center (CePOF)

\end{document}